\documentstyle[11pt,aaspp]{article}
\lefthead{Drukier, G.A.}
\righthead{Models of NGC 6397 -- A. }
\slugcomment{}

\def\Msolar{\ifmmode M_{\mathord\odot}\else$M_{\mathord\odot}$\fi}
\def\et{et\thinspace al.\ }
\def\chiMF{\ifmmode\chi^2_{MF}\else$\chi^2_{MF}$\fi}
\def\chiSDP{\ifmmode\chi^2_{SDP}\else$\chi^2_{SDP}$\fi}

\def\rlrt{\ifmmode r_l/r_t\else$r_l/r_t$\fi}

\begin{document}

\title{Fokker-Planck models of NGC 6397 -- A. The modeling
\footnote{Postscript figures for this paper are available by anonymous
FTP from ftp.ast.cam.ac.uk in the directory
/pub/drukier or by email to the author at
drukier@mail.ast.cam.ac.uk. This paper has been submitted for publication
in {\it The Astrophysical Journal}.}}
\author{G.A. Drukier}
\affil{Institute of Astronomy, University of Cambridge}
\authoraddr{Institute of Astronomy, University of Cambridge, Madingley
Rd., Cambridge, CB3 0HA, England; Internet: drukier@mail.ast.cam.ac.uk}

\begin{abstract}
This is the first of two papers presenting a detailed examination of
Fokker-Planck models for the globular cluster NGC 6397.  I show that
these models provide a good match to observations of the surface
density profile, mass functions at three radii and the velocity
dispersion profile.  The constraint of requiring the best matches to
the mass functions and surface density profiles to occur simultaneously
defines a surface in an initial parameter space consisting of  the cluster
concentration,  mass, and limiting radius.  I discuss various
techniques for locating this surface and the dependence of the quality
of the matches on the position of the model on the surface, the initial
mass function and the retention rate of neutron stars.  The quality of
the matches are usually strongly related to the age of the models, but
one initial mass function was found for which the quality of the
matches are independent of time.
\end{abstract}


\section{Introduction}
This binary paper is a an outgrowth of
previous studies of the dynamics of the globular cluster M71
(\markcite{Lee, Fahlman, \& Richer 1991};
\markcite{Drukier, Fahlman, \& Richer 1992}, hereafter DFR). In DFR
we attempted to compare detailed Fokker-Planck models with
observations of  M71.  The approach taken there was to use star counts
to measure the surface density profile and mass function of the cluster
and radial velocities to measure the velocity dispersion, and then try
to find a Fokker-Planck model to match the observations.  For M71 no
matching model was found and the nature of the discrepancy suggested
that additional physical processes were required in  the modeling.
One of the main lacks in the DFR models, and in all other detailed
comparisons between Fokker-Planck models and observations of globular
clusters, was the absence of any allowance for the effects of stellar
evolution. The difficulties in the case of M71 left open the question
of whether these Fokker-Planck models were relevant to the question of
globular cluster evolution.

Studies previous to \markcite{DFR} had found models to match
observations, but these were for more limited data sets and for
clusters with power-law cusps (\markcite{Grabhorn \et 1992}).
Considering this success, it seemed natural to pursue the question of
relevance by first adding stellar evolution effects to the model, and
then conducting a detailed  comparison  with a large set
of observations of a cusp cluster, in this case NGC 6397.

Since, as will be demonstrated in these papers, matching models can be
found, there are two perspectives that can be taken. One perspective is
that of the numerical modeler who is concerned with the details of the
modeling and the comparison procedure, the size of the initial parameter
space, and questions of uniqueness. The second perspective is much more
narrowly focused and is concerned with what the models tell us
specifically about the current state of affairs in NGC 6397. In order
to prevent an entangled perspective I have decided to split the
discussion of these two aspects into two separate papers.  In this, the
first,  I will discuss the details of the models used, the fitting
procedure and the general results of the modeling. In particular, I
will discuss in some detail the effects  on the models of changes in
the initial parameters.  Here the NGC 6397 data will be treated as a
guide to the interpretation of the models. In the  second paper
(\markcite{Drukier 1994}, Paper B) I will look at the results from the
other angle by examining the details of the best matching models. The
discussions in the two papers are necessarily intertwined and the
second especially will refer back to results and diagrams in this
paper.  The reader might consider them to be an interacting binary.

As it stood in DFR, the Fokker-Planck code, which is descended from the
orbit-averaged, isotropic Fokker-Planck code of \markcite{Cohn (1980)},
had been extended to include a mass function, a tidal boundary
following the formulation of \markcite{Lee \& Ostriker (1987)}, and a
heating term based on the formation and evolution of binaries formed in
three-body interactions (``three-body binaries''; \markcite{Lee 1987};
\markcite{Lee \et 1991}).  The models used here have been further
extended by introducing the effects of stellar evolution.  In DFR, the
models started with the mass function as it would be after a Hubble
time of stellar evolution. That is, it contained a main sequence
terminating at about $0.8 \Msolar$  and the degenerate remnants of the
higher mass stars. In these models it was assumed that the initial
model was at some stage after the massive stars have evolved and that
the further evolution of the lower mass stars was unimportant. Such an
approach is clearly inconsistent with  models meant to follow the full
evolution of a globular cluster. The models presented here remove this
inconsistency by including stellar evolution and pushing the assumed
starting time much earlier conceptually. The expulsion of the left-over
gas from the star-formation process is neglected.  The details are
discussed in the next section.

What is left out of a model can be as important as what is included.
These models assume spherical symmetry and an isotropic velocity
dispersion. The tidal stripping is idealized by assuming that the tidal
boundary is spherically symmetric with respect to the cluster center
and that the strength of the tidal field is constant.  The constancy of
the tidal field excludes both slow changes and tidal shocks.  A
globular cluster in our galaxy can certainly be expected to suffer
tidal shocks  from passages through the disk, passages near the bulge,
and from giant molecular clouds within the disk.  \markcite{Weinberg
(1994)} has recently shown that shock heating can result in large
amounts of mass loss for clusters such as NGC~6397.  Also excluded are
all effects of binaries except for the few ``virtual'' three-body
binaries used as the energy source. \markcite{Gao et. al (1991)}
included an initial population of  binaries as one component in their
two-component Fokker-Planck models. These delay core collapse and leave
the post-collapse clusters with fairly large core radii of between 1\%
and 4\% their half-mass radii.

In many ways NGC~6397 is a useful foil to M71. Both lie at about the
same distance from the galactic center and the galactic plane, but M71
has the metallicity and kinematics of the disk globular cluster system
while NGC~6397 belongs to the halo population.  NGC~6397 is also more
massive and more centrally concentrated than M71 and is regarded as a
post-core-collapse cluster.  From isochrone fitting the age of NGC~6397
is $16\pm2.5$ Gyr (\markcite{Anthony-Twarog, Twarog, \& Suntzeff
1992}).  As discussed in DFR, the dynamical status of M71, ie. whether
it is in a pre- or post-collapse phase, is unclear since the models
give contradictory indications.  NGC~6397, from its high central
concentration, is   highly evolved dynamically and thus is a good
candidate for comparison.  The star count data for NGC~6397 is that of
\markcite{Drukier \et (1993)} supplemented by the mass function from
\markcite{Fahlman~et~al. (1989)}. The velocity dispersion profile of
\markcite{Meylan \& Mayor (1991)} has also been used.

I will begin with the description of the numerical models paying
special attention to the new feature of stellar evolution.  Section
\ref{sec3} will discuss the method used to compare the models with the
observations. Since minimal post-facto scaling is possible with these
models, the results depend only on the initial parameters and the age
of the model. Section \ref{initial parameters} will  define the initial
parameters used here and \S\ref{guidelines} will give an overview of
the effect varying these has on the resulting model.  In total, over
1000 models went into the results to be presented in this paper.  They
were used to refine the description of the effects of parameter
variation on the resulting model fits.  I define a   model set  by
their initial mass function (IMF) and the choice of tidal radius (see
\S\ref{initial parameters}).  I will look first at the largest of these
model sets, will expand the discussion to include model sets with
different tidal radii, and subsequently different IMFs.  I will
conclude by discussing the implications of these findings for future
comparisons.  A more general conclusion appears at the end of Paper B.

\section{Models}
\label{model details}
With the exception of the inclusion of the effects of stellar
evolution, the code used here is basically the same as that discussed
in DFR.  Briefly, I use the isotropic, orbit-averaged form of the
Fokker-Planck equation, where the distribution function is a function
of energy and stellar mass. The clusters are assumed to be spherically
symmetric.  The coupled Fokker-Planck and Poisson
equations are solved by the two step process discussed more fully in
\markcite{Cohn (1980)}.  First the diffusion coefficients are
calculated and the distribution function is advanced in time in
accordance with the Fokker-Planck equation.  At this point the
potential and the distribution functions are no longer consistent, so
the second step is to solve the Poisson equation subject to the
constraint that the distribution function remains the same function of
the adiabatic invariant $q(E)$ in the notation of
\markcite{Cohn(1980)}.  The solution of the Poisson equation is done
iteratively.

In order to reverse core collapse, an energy source is required.  Here,
I estimate statistically the number of binaries formed in three-body
encounters and the energy released by each such binary as it is
hardened by interactions with the field stars.  At any time there are
only a few such binaries present in the cluster, which is why the
treatment is statistical.  The prescription for doing this is discussed
in \markcite{Lee \et (1991)} and \markcite{DFR}. Alternative sources of
energy are from initial population of binaries, or from binaries
formed by close encounters and subsequent dissipation of orbital energy via
tides in their atmospheres.  These processes are not included in the
models presented here.

The tidal boundary is imposed in energy space by defining the
tidal energy boundary as the potential at the radius enclosing a fixed
mean density. This radius is referred to as the tidal radius $r_t$. The
distribution functions are reduced exponentially for energies beyond
this boundary with the stripping rate dependent on the difference
between the energy and the tidal-energy.  The rate is given by the
formula in \markcite{Lee \et (1991)} based on the derivation of
\markcite{Lee \& Ostriker (1987)}.  Two modifications have been made
here.  The first is to  remove  the discontinuity in the first
derivative of the tidal stripping rate with respect to energy. Since
the distribution function is a function of energy and not the adiabatic
invariant, the iterative solution of the Poisson equation requires that
it be regridded in $E$ to preserve the dependence on $q(E)$.  The
regridding is done via a second-order Taylor expansion of the the
distribution function in terms of the adiabatic invariant. The effect
of the discontinuity in the stripping rate is to introduce
discontinuities in the first derivative of the distribution function.
These are then amplified by the regridding procedure and can lead to a
catastrophic failure of the Poisson equation solver.  To reduce  the
chances for this, the  stripping rate has been smoothed over the
transition region using a cubic polynomial  which is required to be
continuous to  the first derivative. This is described in Appendix
\ref{app:A}

The second modification concerns the timing of the tidal stripping
phase with respect to the two stages of advancing the Fokker-Planck
equation and updating the potential.  Since we need both the density
profile and the  potential to define the tidal boundary, the tidal
stripping must be done after the potential is updated and is once again
consistent with the distribution functions, but before proceeding with
the next Fokker-Planck step.  This was the method used by \markcite{Lee
\& Ostriker (1987)}.  The problem with this is that the stripping is
done on the distribution functions and afterward the  potential and
densities are once again inconsistent with them.  If the amount of
tidal stripping is small, this is only a small inconsistency; but in
the late stages of the model's evolution the mass becomes small and the
tidal losses proportionately large.  The mass decreases approximately
linearly with time and since  $r_t\propto M^{1/3}$, $\dot{r_t}\propto
M^{-2/3}$.  In cases where the rate of decrease in $r_t$ is large it
becomes necessary to find a self-consistent solution.  To correct for
this problem an iterative scheme for simultaneously doing the tidal
stripping and solving the Poisson equation was used.  This scheme is
described in Appendix \ref{app:B}

In their Fokker-Planck model, \markcite{Chernoff \& Weinberg (1990)}
used a somewhat different scheme for tidal stripping. They found that
in the late stages they were  unable to find a self-consistent solution
to the Poisson equation and the tidal boundary condition. What I find
using the iterative scheme is that a self-consistent solution was
possible in these situations. The difference arises in that the tidal
boundary condition of \markcite{Chernoff \& Weinberg (1990)} is
equivalent to $f(E)=0$ for $E>E_t$ (as defined in Appendix \ref{app:A})
which is discontinuous at $E=E_t$.  The tidal stripping condition  used
here ensures continuity in $f(E)$ and its first two derivatives.  Thus,
a self-consistent solution is still possible even with extreme rates of
tidal mass loss.

The addition of the effects of stellar evolution is the main change in
the code from DFR.  The approach used here is much
the same as that used in \markcite{Chernoff \& Weinberg (1990)}. Even
without stellar evolution a mass spectrum is desirable.  To introduce
this a mass grid is employed, with the mass spectrum being broken into
a series of bins, each with its own mass and distribution function.
Each of these can be considered  a mass species which is meant to
represent a range of stars with similar mass.  The initial mass for
each mass species is taken to be  the geometric mean of the masses at
the bin boundaries.  In order to account for stellar evolution, we
simply allow the mass for each mass species to change with time.  The
simplest way to specify this is to adopt functions for the
stellar lifetimes and final masses for stars of a given initial mass.
I assume that the stars evolve instantaneously from their initial
masses to their final masses without worrying about the details of
stellar evolution.

To be more specific, let $t(m^i)$ be the lifetimes of stars with
initial mass $m^i$ and let $m^f(m^i)$ be their final masses.  I assume
that at time  $t(m^i)$ a star with initial mass $m^i$ becomes a star
with mass $m^f(m^i)$. The initial mass function (IMF) is given by
$N(m^i) dm^i$.  This is often taken to be a power law
\begin{equation}
N(m)\,dm \propto m^{-(x+1)}\,dm.
\end{equation}
When the mass spectral
index (MSI), $x$, is defined this way, the Salpeter mass function has
$x=1.35$.  For a bin $j$ with boundaries $m_{j-1}^i$ and $m_j^i$,
$m_{j-1}^i < m_j^i$, the total mass in the bin is given by
\begin{equation}
 M_j=\int_{m_{j-1}^i}^{m_j^i} m^i N(m^i)\, dm^i.
\end{equation}
The initial  mass of mass species $j$ is taken to be
$\overline{m}^i_j = \sqrt{m_{j-1}^i m_j^i}$ and the initial number of
stars in the bin is $N_j=M_j/\overline{m}^i_j$. The final mass for mass
species $j$ is $ \overline{m}^f_j = m^f(\overline{m}^i_j)$. The mass of
species $j$ is assumed to change linearly from $\overline{m}^i_j$ to
$\overline{m}^f_j$ over the time interval $t(m_j^i)$ to
$t(m_{j-1}^i)$.  Since $t(m^i)$ is a monotonically increasing function,
$t(m_j^i) < t(m_{j-1}^i)$.

The main effect of mass loss due to stellar evolution is to reduce the
depth of the potential and indirectly ``heat'' the cluster.  To remain
in virial equilibrium in the shallower potential, the kinetic energy of
the cluster stars must also be reduced.  Due to the negative specific
heat of self-gravitating systems, this results in a net expansion,
especially in the core. The closer to the cluster center the mass loss
takes place, the more effective is the heating and the stronger the
expansion.  If, as is done here, the model starts with  the relative
proportions of the various mass species  the same at all radii, then
the effectiveness of stellar evolution in causing the expansion will
depend on the ratio of the stellar evolution time scale, $t_{se}$, to
the dynamical evolution time scale.  If $t_{se}$ is long compared to,
for example, the central relaxation time, then the most massive stars
will have time to sink to the center of the cluster through dynamical
friction before they evolve. Their evolution is then a more effective energy
source.  The practical effect of the stellar evolution mass loss is to
expand the cluster and delay core collapse beyond the time expected
without such mass loss.  The length of the delay, if indeed the mass
loss doesn't destroy the cluster entirely, is strongly dependent on the
IMF, $m^f(m^i)$, $t(m^i)$, and the initial structure.

\markcite{Chernoff \& Weinberg (1990)} based their  stellar lifetimes
on the \markcite{Miller \& Scalo (1979)} compilation for Population I
stars. For $m^i<4.7\Msolar$  the masses of their white dwarf remnants
were based on the formula of \markcite{Iben \& Renzini (1983)} with $
\eta =\onethird $.  The intermediate mass stars ($4.7<M/\Msolar<8$)
were assumed to be completely destroyed in a supernova. Stars with
$M>8\Msolar$ were assumed to leave a 1.4\Msolar\ neutron star. For
purposes of comparison I ran a model corresponding to a model with a
\markcite{King (1966)} model dimensionless central potential $W_0=7$
and $x=1.5$ in family 3 of \markcite{Chernoff \& Weinberg}. The results
of the two models were very similar once differences in the choice of
Coulomb logarithm were taken into account.

In terms of finding a model to match NGC~6397, the stellar lifetimes
chosen by \markcite{Chernoff \& Weinberg (1990)} are not useful because
of the effects of metallicity on stellar lifetimes.  What is needed are
the lifetimes of stars of all masses for $[Fe/H]=-1.9$ appropriate to
NGC~6397.   Stellar models of the appropriate metallicity are available
for low-mass stars, but no such models have been published for masses
above about 0.95 \Msolar.  The most extensive set of stellar evolution
models are those by the Geneva Observatory group (\markcite{Schaller
\et 1992}; \markcite{Shaerer \et 1993a,b}; \markcite{Charbonnel \et
1993)}.  These cover the mass range from 0.8 to 120 \Msolar, but only
extend to $Z=0.001$.  For stars with the metallicity of NGC~6397,
models with $Z=0.0002$ are required.  To estimate these, the lifetimes
until the end of He burning for stars of varying metallicity at
constant mass were taken from the Geneva models.  For $Z\le 0.008$ the
lifetimes of stars with $m>5\Msolar$ is approximately constant with
varying metallicity so the lifetimes for  $Z=0.001$ were adopted.  For
$m<5\Msolar$ the lifetimes were extrapolated to $Z=0.0002$ by a
polynomial fit to the lifetimes for $Z\le 0.008$.  For $m<1.25 \Msolar$
the lifetime goes as $m^{-3.5}$ and lifetimes for stars with $m<0.8
\Msolar$ have been extrapolated assuming this power law. As a
consistency check, I compared these ages with ages derived from the
models of \markcite{VandenBerg (1992)}.  For the [Fe/H]$=-2.030$
models, the ages agree to within 5\%. The lifetimes are given in Table
\ref{ages}.

There are several options for the choice of $m^f(m^i)$.  First,
for $M>8\Msolar $ I have assumed that the remnants are 1.4\Msolar\
neutron stars.  The situation for the stars which become white dwarfs
is more complicated. Figure 3 of \markcite{Weidemann and Yuan (1989)}
shows over a dozen different proposals and models for this relation
with a wide range of properties. The effect of varying the $m^f(m^i)$
relation is to change the total amount of mass lost from the cluster
through stellar evolution and to change the rate at which the mass is
lost.  This is  most important in the early stages. A
higher mass loss rate and a larger total mass loss at the same rate
results in a greater expansion of the cluster.  The effect on the time
of core-collapse is more complicated. While a large cumulative mass
loss results in a more expanded cluster, the total mass is also
smaller. This reduces the relaxation time and could cause a quicker
collapse.  For the models described here I have used the scheme of
\markcite{Wood (1992)}:
\begin{equation}
m^f(m^i)= 0.4 e^{m^i/8.}
\label{mfmi equation}
\end{equation}
for $0.5\Msolar<m^i<8.0\Msolar$. Since stars with
masses less than 0.5 \Msolar\ do not evolve until long past the epoch
we are interested in, we need not worry about their final masses.

All of the models I discuss in these papers have $t(m^i)$ based on
Table \ref{ages} and $m^f(m^i)$ using eq. (\ref{mfmi equation}) for the
stars less massive than 8\Msolar\ and 1.4\Msolar\ for the  more massive
stars.

\section{Finding a match}
\label{sec3}

Locating a matching model for a particular set of observations is not
an easy task. The available parameter space is large and the effects on
the resulting models  of changing individual parameters  are complex
and non-linear.  The interaction of the various parameters provide
tradeoffs which can be played against each other to achieve the desired
end, but can also lead to models with quite different initial
conditions leading to equally good matches.  Experience does lead to
some useful guidelines and these will be discussed in
\S\ref{guidelines}

\subsection{Comparison procedure}
\label{comparison}
To tell how well a particular model matches the NGC 6397, the
results of the model must be compared with the observations. Once a set of
initial parameters has been decided on, the model is run and its state
is periodically saved. Of these data, I have taken those in
the interval between 10.5 and 19 Gyr for further analysis.  In terms of
scaling, once the tidal boundary, the three-body reheating mechanism
and stellar evolution have been included in the model there are no
global scales left  free for adjusting.  Thus the comparison procedure
is fairly straight forward.  The data available for comparison with the
models are the surface density profile (SDP) and two mass functions (MFs) from
\markcite{Drukier \et (1993)}, the intermediate-distance mass-function
from \markcite{Fahlman~\et (1989)} and the velocity dispersion profile
from \markcite{Meylan \& Mayor (1991)}. I will follow the naming
convention from \markcite{Drukier \et (1993)} and refer to the three
mass functions as the du~Pont:if, FRST, and du~Pont:out MFs in order of
distance from the cluster center.

Since the observations are based on the projected distribution of stars
in the cluster, the first thing to do is to project the density
distributions and velocity dispersions for each mass species in the
model.  I then simulated the observing procedure by  integrating the
projected profiles over appropriate regions.  For the mass
functions, the projected densities for each unevolved mass species were
integrated over rectangular regions with the same size and orientation
and at the same radial position as the observed fields.  The widths of
the mass bins were then used to convert the integrated counts into
numbers per unit mass.  The observed  and model mass functions are not
on the same grid, so the model mass function is interpolated to give
values at the observed mean masses.  Since the model MF is smooth this
is not difficult.  A $\chi^2$ statistic is calculated for each of the
three pairs of observed and model mass functions using the
observational uncertainties as weights.  The quality of the match is
judged by the mean of the three mass function $\chi^2$ statistics,
\chiMF.

Sets of annuli were defined matching the observed radii and mean
densities within the annuli were integrated from the model density
profile.  There is a slight inconsistency here in that many of the
observed data points are from sections of annuli rather than full
annuli. The mean of $\chi^2$ from the two magnitude limited surface
density profiles, \chiSDP, was used as the figure of merit for the SDP
fits.

There are several issues to be addressed before proceeding with the
comparison of the model and observed profiles.  First, the observed
profile is for stars above the main-sequence turn-off. This is because
of the brightness of these stars and the high degree of crowding in the
images of this concentrated cluster. Therefore, the mass bin to use for
comparison is the one which is currently evolving.  With the scheme
used for implementing stellar evolution, the mass of the currently
evolving bin is usually less than the mass of the stars at the
turn-off.  Once the mass drops, the stars in the evolving mass species
becomes less concentrated and the mass species is no longer suitable
for comparison.  Instead of the evolving species I have used the next
less massive one.  The interval between 0.74 and 0.90 \Msolar\ has been
divided into seven mass species to ensure that the mass discrepancy is
small.\footnote{ For technical reasons, the first grid of models was
run twice, with slightly different sets of mass bins.   In
Figs.~\protect{\ref{rlrt}}--\protect{\ref{chi contour}} I will show
some contour diagrams giving the results for the binning discussed in
the previous paragraph.  There was an earlier set of models which only
had the finer coverage for five bins with masses from 0.77 to 0.84
\Msolar.  (The range was increased to ensure that the mass of the next
bin would be close to the turn-off mass for models with ages from 10.3
to 18.5 Gyr.) For both grids of models, the contour plots show the same
large-scale  features. This demonstrates that the results do not have a
strong dependence on the choice of mass bins.}

The second issue relates to the widths of the mass bins and the range
of masses in the observed surface density profiles. \markcite{Drukier
\et (1993)} produced two surface density profiles with different
magnitude limits.  The shallower profile ($I<14.$) extends into the
center of the cluster while the deeper profile ($I<15.5$) is limited to
radii greater than 20\arcsec\ from the cluster center.  Without
detailed information on the mass-luminosity relationship for the
observed stars it is very difficult to measure the mass range
observed.  Therefore, one free scaling parameter was allowed for in comparing
the model and observed surface density profiles.  When the
\chiSDP\ statistic was calculated a single rescaling was also fit by
minimizing the $\chi^2$ for each of the two SDPs.  The two profiles
were fit separately and the known offset of a factor of 3.09
(\markcite{Drukier \et 1993}) was employed.  The mean rescaling was
adopted and \chiSDP\ calculated.

Similar techniques could not be employed for the velocity dispersions
since the observed velocities were not available.  Rather, the
projected velocity dispersion profile for the model was plotted
together with the observed data points and used to confirm that the
time which best fit the surface density profile and the mass functions
also matched the dynamical information.

It should be kept in mind during the comparisons discussed below that
\chiMF\ and \chiSDP\ are independent  estimators of the quality of
fit.  The optimal model will be one that minimizes both at the same
time and which also gives a velocity dispersion profile consistent with
the data of \markcite{Meylan \& Mayor (1991)}. The age of the model at
the optimal time should also be the age of stars in the cluster.  Given
the large uncertainties in determinations of the absolute ages of
globular clusters, this requirement will not be applied too strictly,
but the age should be between 13 and 18 Gyr.  As will be seen, once a
range of parameters giving good matches is found, locating the best of
these becomes a fine tuning problem.  In general, I have not tried to
find the best-matching age for any given model, but have just adopted
the best of the model dumps.  While the model could be rerun with
finer time resolution, the differences in the
fits are small enough to be unimportant given the quality of the data.
The models shown in Paper~B have been rerun this way
at around the age of their best match.

\subsection{Initial parameters}
\label{initial parameters}
The IMF I
initially used was the same as IMF J in \markcite{Drukier (1992)}
where, of the 10 tried, it provided the best fit to the observed
NGC~6397 mass functions.  In \S\ref{variations}  I will discuss the
effects of variations in the IMF.  The mass gridding has been changed
to allow for finer gridding between 0.74 and 0.90 \Msolar\ as discussed
above.  The IMF is made up of two power-laws, one with mass spectral
index 1.5 for $m<0.4\Msolar$ and the second with $x=0.9$ for
$m>0.4\Msolar$.  The relative scalings were set so that the mass
function is continuous at 0.4 \Msolar. I will refer to models made with
this IMF and the set of stellar data in \S\ref{model details} as
``U20'' models.

The models start as \markcite{King (1966)} models with all species
having the same initial profile.  The initial structure of the model is
defined by four parameters. The first is the dimensionless central
potential of the King model, $W_0$.  The strength of the tidal field is
given by the initial tidal radius, $r_t$ and the initial limiting
radius of the model is $r_l$.  A more useful way to parameterize this
is to use the ratio $r_l/r_t$.  If ratio is unity  then the initial
model fills its tidal volume, but this need not be the case.  If
$r_l/r_t <1.$ then the model has room to expand before suffering
substantial tidal losses. The models are taken to travel on a circular
orbit so that the strength of the tidal field is constant. The
treatment of tidal stripping is further limited by the assumptions of
spherical symmetry and an isotropy velocity dispersion.  The tidal
radius, $r_t$, is given in terms of a cluster with mass $10^5
\Msolar$.  The initial mass, $M_0$, completes the specification of the
model once $r_t$ is rescaled by $\left(M_0\over
10^5\Msolar\right)^{1/3}$ and the initial limiting radius calculated
from $r_l/r_t$. Unless otherwise specified, I will give $r_t$ as the
value for a $10^5\Msolar$ cluster.  In this way a value of $r_t$ can
be thought of as specifying the galactocentric distance of the model,
$R_G$,  by assuming a galactic mass model, $M_G(R_G)$, and taking $M_c
= 10^5\Msolar$ in  the equation
\begin{equation}
r_t={2\over3}{\left[M_c\over{2M_G(R_G)}\right]}^{\onethird}  R_G.
\label{tidal equation}
\end{equation}

\subsection{Guidelines}
\label{guidelines}
As can be appreciated from
the preceding discussion, the available parameter space is large and a
systematic approach is required. To begin with I searched through the
$(W_0,M_0,r_t,\rlrt)$ parameter space for an acceptable model keeping
the following guidelines and their converses in mind. I refer to this
searching stage as the ``hunting'' mode of running models.

\begin{enumerate}
\item In general, an increase in the
relaxation time  will cause a later core collapse.  It also
decreases the rate of mass loss through the tidal
boundary.
\item Increasing the mass or the limiting
radius or reducing $W_0$ will increase the
relaxation time. (An increase in $W_0$ does not necessarily  lead to an
earlier core collapse however.  If $W_0$ is large enough, a further
increase, by  decreasing the central relaxation time, increases the
initial mass segregation and the depth of the central potential and
increases the amount of expansion due to stellar evolution
for the same amount of mass loss.  The maximum size reached in the
expansion phase can then be larger for a larger  $W_0$ and core
collapse is  correspondingly delayed.)
\item As $r_l/r_t$ decreases
the tidal mass-loss rate decreases. Thus a decrease in $r_l/r_t$ gives
the result that at any given time the mass and the relaxation time are
increased, all else being equal.
\end{enumerate}

Unfortunately, it is difficult to quantify any of these effects as they
are also strongly dependent on the IMF. They can be traded off against
one another in finding a better fit.  As discussed below, the fits to
the mass functions and the surface density profile both show well
defined minima as a function of age.  What I call a ``well-fitting
model'' will be one where both these minima occur simultaneously.
The minima in \chiSDP\  occur close to, but before, core collapse, so
the time of core collapse is a useful marker. The minima in
\chiMF\ cluster around a optimal mass (see \S 2.1. in Paper~B) and
can be thought of as occurring when the model mass reaches this value.
In Fig.~\ref{chi curves} I show  \chiMF\ and \chiSDP\ as a function of
time for  one model. The two minima are not aligned and it is desirable
to change the parameters to bring them into alignment at, preferably,  an age
older than 13 Gyr.  To achieve this the following rules came in
handy for this data set:
\begin{itemize}
\item Changing
$M_0$ alone tends to move both the time of optimal mass and the time of
core collapse by about the same amount.
\item Changing $W_0$ alone
tends to not affect the time of  core collapse by very much, but does
change the time of optimal mass.
\item Increasing $r_l$ alone (ie. \rlrt) reduces
the time of optimal mass and, to a lesser extent,  the time of core
collapse.
\end{itemize}
The inter-relationship of these rules defines the parameter surface
containing the good models.  For other sets of observations and in
other clusters different relationships may apply. In each case, the
sensitivity of the results to changes in the initial parameters need to
be estimated. They can then be used to formulate similar rules
applicable to those data.

\section{Results}
\label{Results}
\subsection{The Parameter Surface}
\label{surface}
In order to pursue the idea of a lower-dimension surface defined by the
well-fitting models I ran a grid of models in the three-dimensional
space defined by $W_0$, $M_0$, and \rlrt. As I explain in Paper B, the
optimal value for $r_t$ is around 20 pc, but there is a wide range of
acceptable values.  The well-fitting models found in the hunting stage
had $r_t = 18 $ or 19~pc so it was more straight-forward to look for a
surface with $r_t = 18.5$ since points on the surface were already
approximately known.  I later ran model sets with $r_t=17$, 20, and 21
pc and will discuss them further below.  For now, it suffices to note
that the choice of $r_t$ does not affect the results very strongly. The
region covered was $4.01<W_0<6.47$, $3.41<M_0<9.94$, and
$0.4<\rlrt<1.10$.

In Fig.~\ref{chi curves} I show  \chiMF\ and \chiSDP\ as a function of
time for a typical model.  Note that this is not what I have been
calling a ``well-fitting'' model since the two minima are not
coincident.  Defining the difference in the time of minima  as $\Delta
t \equiv ($time in minimum in $\chiMF)  - ($ time of minimum in
$\chiSDP)$, the locus of well-fitting models is that region of
parameter space where $\Delta t = 0$.  As might be expected,  the
well-fitting models define a surface in $(W_0,M_0,\rlrt)$ space.  Since
only a fraction of the models in the grid happen to lie on this
surface, I estimated the position of the surface by interpolating
along the grid.  I took all the pairs of models differing in only one
parameter and with $\Delta t$ of opposite signs and used linear
interpolation between them  to find the third parameter. The same
procedure for these models, together with the data from the  models
with $\Delta t=0$, gave the age of the model, \chiMF, and \chiSDP\  on
the $\Delta t =0$ surface. This surface is fairly smooth, but has some
thickness ($\pm 0.4$  Gyr) due to the finite time resolution in the model
results (see. \S\ref{Discussion}).

Figure~\ref{rlrt} shows this surface of well-fitting models.  The
contours give the estimated value of \rlrt\ as a function of $W_0$ and
$M_0$.  The squares indicate the positions of the estimates used in
constructing the contours and the  circled squares are models which had
$\Delta t=0$.  Contouring algorithms generally require points on a
regular grid, so for the contour diagrams I defined a grid in
$\left(W_0,M_0\right)$ and used bi-linear interpolation to estimate the
desired datum at each grid point from the values at the three nearest
data points. Any grid point which did not have three data points closer
than 3.5 grid spacings were ignored.  These ignored points are
indicated by dots in Fig.~\ref{rlrt} and show the size of the
interpolating grid. Features on this scale are artifacts of the
contouring process.   The contours are spaced by 0.05 in \rlrt\ with
thicker contours every 0.25.  The value of \rlrt\ increases from
lower-left to upper-right with the thick contour on the right side
being $\rlrt=1$.  Contours in this diagram were a very
reliable guide in  estimating the value of \rlrt\ as a function of the
other two parameters.

Figure~\ref{time} shows the age of the models on the  $\Delta t=0$
surface. The contours are at 1 Gyr intervals with the thick lines
indicating 12 and 15 Gyr.  Models with ages less than 10.5 Gyr and
greater than 19 Gyr have been excluded from these  contour plots.  The
upper age limit defines the top-left edge of the contoured region, the
lower limit, the lower-edge.  Additional good models with higher
concentrations probably exist, but some tests with $W_0=9$ models with
this IMF indicate that these all core collapse very quickly. An
extrapolation from Fig.~\ref{rlrt} suggests that high-concentration
models would also need to have values of \rlrt\ much larger than one,
ie. we would have to assume that globular clusters start with sizes
much larger than the tidal limit imposed at the galactocentric distance
of their origin. This is not an unreasonable suggestion.  For the high
concentrations being considered (eg. $W_0 > 6$), most of the mass is
well within the initial tidal boundary.  Further, contrary to the
assumption here, real globular clusters travel on eccentric orbits and
thus feel a time-varying tidal force.  If a globular cluster moved
closer to the center of the galaxy after its birth, then it would, in
effect, be starting its evolution overflowing its tidal boundary. Such
high $W_0$ models will not be discussed here.

Figure~\ref{chi contour} show similar contour plots for   (a) \chiMF,
(b) \chiSDP\ and (c) their mean.  There are several features to note in
these diagrams.  First, there is a broad region in Fig.~\ref{chi
contour}a where the models match the observed mass functions.  The fit
improves with higher initial mass at a given initial concentration and
the dependence on initial concentration is weak. The models with the
lowest initial masses evolve fairly quickly and, if not already
excluded for being younger than 10 Gyr, would be excluded for giving a
poor match to the MFs.  The fits to the mass functions are quite
satisfactory in much of the parameter space.

The match for the surface density profile is more problematic.  A
comparison of Fig.~\ref{chi contour}b with Fig.~\ref{time} shows that
the contours of constant \chiSDP\ are parallel to the contours of
constant age.  Further, \chiSDP\ increases with age and is less than
two only for models younger than 14 Gyr.  Figure~\ref{chi contour}c
shows the contours of constant $(\chiSDP+\chiMF)/2$ which I use as an
overall figure of merit.  This mean $\chi^2$ is dominated by \chiSDP,
but at young ages the increase in \chiMF\ becomes important. The result
is a valley in the mean $\chi^2$ contours where the best models lie.
The lower boundary of the valley is truncated in the contour plot by
the lower age cutoff in the models, but is readily apparent in the
original numbers. This region is occupied by models with ages between
11 and 13 Gyr.

The dependencies of \chiMF\ and \chiSDP\ on time implied in
Fig.~\ref{time} and \ref{chi contour}a and b are shown more explicitly
in Fig.~\ref{chi vs. t}. This plots \chiMF\ and \chiSDP\ against the
age of the model for all the points defining the $\Delta t=0$ surface.
Clearly, the best fitting  models have an age of about 12 Gyr. If it is
assumed that the U20 IMF is correct, then these models would imply that
NGC~6397 is 12 Gyr old.  This age does contradict the age derived from
isochrone fitting ($16\pm 2.5$ Gyr, Anthony-Twarog, Twarog, \& Suntzeff
1992) and, if it were correct, would suggest a problem either with the
stellar evolution models, or with these dynamical models.  However,
models run with other IMFs (see \S\ref{variations}) give either older
ages, or no preferred  age for NGC~6397.  In view of this, the safest
thing to do is to reject the assumption that the U20 IMF is correct.

Why is there such a strong dependence on the fit of the SDP with time?
As discussed in \markcite{Cohn (1985)} and \markcite{Chernoff \&
Weinberg (1990)} the  central density profile for stars with mass $m_k$
in this sort of  model will have a logarithmic slope
\begin{equation}
\zeta_k = -{d \ln \rho_k\over d \ln r}= \left(1.89{m_k\over m_u}+0.35\right),
\label{slope}
\end{equation}
where $m_u$ is the mass of the species
dominating the core.  Projection effects make the observed surface
density flatter by one.   Clearly, as the model ages the mass of the
stars at the turn-off decreases and it is these stars which are counted
for the surface density profile.  Given  a core dominated by 1.4\Msolar\
neutron stars, the \chiSDP\ result requires that the turnoff stars be
more massive than 0.83\Msolar.  The correlation of \chiSDP\ with time
is a reflection of the dependence of turn-off mass on the age of the model.
The correlation would not be changed if a different $t(m^i)$
relation were used. The only difference would be in the initial parameters
needed to produce a well-fitting model of a desired age.

When models are run with a different choice of $r_t$ the same principle
still applies.  In Fig.~\ref{rt=}(a) to (c) I show contour plots of
\rlrt, $t$ and the mean $\chi^2$ for a U20 model set with $r_t=20$ pc.
The parameter surface is very similar to that for $r_t=18.5$ pc, but
the models are about 1.6 Gyr older at a given point on the surface.
The lines of constant \chiSDP\ are shifted by a similar amount, but
retain the same relationship to the age of the model.  Figure~\ref{rt=}
(d) to (f) shows a similar series of contour diagrams for $r_t=17$ pc.
In this case the shift is in the opposite sense, with the models being
about 1.6 Gyr younger than the $r_t = 18.5$ pc models at a given place
on the parameter surface.  Again, the \chiSDP\ contours shift with the
age contours.

Column 3 in Fig.~\ref{grids} show the time dependencies of
\chiSDP\ and \chiMF\ for four sets of models with IMF U20 and $r_t$ as
indicated in the right margin. The $r_t=18.5$ pc panel in this column
is based on  Fig.~\ref{chi vs. t}.  (The other of model sets
shown in this figure will be discussed in the next section.) The
distribution of points with time is a result of the varying coverage of
parameter space for each set of models. All four model sets display
very similar temporal dependencies although the initial parameters
giving rise to a point with a given age are different for each model
set.

Since all the well-fitting models have much the same structure in terms
of the mass and half-mass radius, it is not surprising that there is
very little difference between them in terms of half-mass relaxation
time.  The number of elapsed half-mass relaxation times,
$\tau\equiv\int{dt\over t_r}$, also shows little variation amongst all
the well-fitting U20 models. Within any set of models $\tau$ increases
by about 15\% between 12 and 18  Gyr.
Hence, the differences in the SDP matches are not a result of the
older models also being significantly more evolved dynamically.  This
may be reflected in the small differences in the model mass functions,
however.

\subsection{Uniqueness}
\label{variations}
So far, we have seen that
the observations cannot uniquely constrain the initial parameters for
NGC~6397, but only a subset of them.  Can they constrain the IMF and
other stellar data? I have addressed this question by constructing five
additional sets of stellar data and then searching for well-fitting
models with these sets of data.  All of these models retain the same
IMF for the mass range 0.1 to 2.0 \Msolar.

One of the five schemes, the ``NNS'' models assumes that all neutron
stars receive a sufficiently large ``kick'' velocity at their birth to
escape the cluster. (Field pulsars are known to have high space
velocities [\markcite{Gunn \& Ostriker 1970}]. Current estimates
suggest that the mean velocity is 450 km s$^{-1}$ [\markcite{Lyne \&
Lorimer 1994}]. Whether these originate due to asymmetries in the
collapse or result from the unbinding of binaries is still open to
question [qv.  \markcite{Wijers \et 1992}, \markcite{Bailes 1989}].)
This model set serves as the alternative limiting case to the retention
of all the neutron stars. The other four schemes involve
modifications to the IMF.  Three change  the mass range by extending
the lower limit to 0.05 \Msolar (the ``L05'' models), extending the
upper limit to 30 \Msolar (the `` U30'' models), and restricting the
upper limit to 10 \Msolar (the ``U10'' models).  The fifth set of
models, the ``X2'' models have second break in the IMF, this at
2\Msolar, and a mass spectral index $x=2$ for more massive stars.
(Recently \markcite{Hill, Madore \& Freeman (1994)}  found a MSI $x=2\pm
.5$ for $m>9\Msolar $ in a selection of Magellanic Cloud associations.
Their study also suggested a somewhat smaller MSI for lower mass stars.)
The parameter space was searched in both the hunting mode and by more
systematic searches along a  grid in the $(r_t=18.5{\rm
\ pc},W_0, M_0, \rlrt)$ parameter space.  U10 and X2 model sets were also
run with $r_t=20$ pc.  Table~\ref{IMFs} gives the mass fractions for
the various mass components in each IMF. Note that the IMFs
are the same in the NNS and U20 IMFs, but that the
high mass stars leave no remnants.

For the X2 and NNS models no satisfactory matches were found.  In all
cases where the models had evolved through core collapse, minima in
\chiSDP\ were seen both before and after core collapse, with a maximum
at core collapse.  In these models, the mass of the evolving stars is
much closer to the mean mass in the core and therefor they show very
steep surface density profiles at the time of core collapse (see
Fig.~12 in Paper~B).  The surface density drops off more quickly
than observed in the outer region covered by the mass functions.  The
segregation measure $S_r($m;du Pont:out,du Pont:if) (defined by eq. (2)
in \markcite{Drukier \et 1993} as the logarithm of the ratio of the two
mass functions) between the du Pont:if and du Pont:out mass functions
is much larger than observed indicating that these models suffer too
much mass segregation.  Coincidences between local minima in
\chiMF\ and \chiSDP\ occurred in both the collapsing and post-collapse
phases.  The collapsing models still have large core radii and gave
quite poor matches to the surface density profile.  For the X2 models,
the mass functions were matched better when $r_t=20$ pc was used, but
there was no improvement to the match of the SDP.  Post-core-collapse
models have profiles that match the observed SDP fairly well, but these
are all older than 18.5 Gyr and the matches to the mass functions are
very poor.

Figure~\ref{grids}  summarizes the time dependence of \chiMF\ and
\chiSDP\ for the model sets with well-fitting models.  The U30 model
set has much the same time dependency as do the U20 model sets.  The
best U30 models lie at a somewhat younger age than do the U20 models.
This is understandable in terms of the argument given above since the
U30 models have a higher proportion of heavy remnants than do the U20
models and the effective $m_u$ in equation (\ref{slope}) is higher.
For the same observed slope at core collapse, the mass of the turn-off
stars must be higher and thus younger. The effect is small, but, given
the even larger contradiction between the optimal age here, and the
isochrone  age of NGC~6397,  a higher number of neutron stars can be
excluded. Panels (a) to (c)  of Fig.~\ref{LRTt} shows \rlrt, the mean
$\chi^2$, and age of the models on the $\Delta t=0$ surface for the U30
model set. The behavior is very similar to that of the U20 model sets.
Note that the shapes of the contoured regions are determined by the
range of models runs.  There certainly exist well-fitting models beyond
these regions, I just haven't looked for them.

That the L05 model set has a later time when the models best fit all
the observations is not surprising given the comparison between the U20
model sets and the U30 model set. On the other hand,
the dependence of \chiMF\ on time is much stronger than for
the U30 and U20 model sets. It is as strong,
though in the opposite sense, as the dependence of \chiSDP\ on time.
The mean $\chi^2$ is fairly constant with time, but at about 14 Gyr the
trade-off between the two is minimized. Panels (d) to (f) of Fig.~\ref{LRTt}
confirms that the dependence of the mean $\chi^2$ on time is much
weaker than for the U20 or U30 models. There is a large basin of models
with $W_0$ between 4.5 and 6.0 and $M_0$ between $4\times 10^5$ and
$6\times 10^5\Msolar$ which give matches of similar overall quality by
trading off between \chiMF\ and \chiSDP.

The behavior of the U10 models serves as a warning against
extrapolation.  For $r_t=18.5$ neither $\chi^2$ shows any time
dependence at all.  Further, the mean $\chi^2$ for this model set is
the lowest of any. \chiSDP\ is at all times as low as that seen in
any other model and \chiMF\ although globally higher than in some other
cases, is no higher than its value at the time of minimum mean $\chi^2$
in the other model sets.  An additional model set was run with IMF U10
and $r_t=20$ pc.  In this  model set \chiSDP\ is still approximately
constant with time, and \chiMF\ now decreases with time; the minimum in
the mean is at over 18 Gyr.  The difference in the time dependence of
\chiMF\ between the $r_t=18.5$ pc and $r_t=20$ pc models is consistent
with a trend to steeper slope in \chiMF\ vs. $r_t$  in the U20 models.
Panels (g) to (l) in Fig.~\ref{LRTt} make clear the very weak
time dependence of the mean $\chi^2$.

To further investigate this problem, Fig.~\ref{MF lines} presents
$\chi^2$ for each of the three mass functions separately as a function
of time for each of the model sets. To reduce the clutter in the
diagram I have just plotted the best fitting straight line through each
of the sets of estimates.
Systematic trends are visible in the U20 column for both the slopes
and intercepts of the $\chi^2$ lines and these trends are also
present in the two  U10 model sets. The  decrease in
$\chi^2$ for the du Pont:out MF with increasing $r_t$ is
understandable as indicating that larger tidal radii are preferred in
matching this region of the cluster.  On the other hand, the fit to the
du Pont:if MF gets worse as $r_t$ is increased. To expand on the points
made at the end of \S\ref{surface}, the variation in $\tau$  in any
single model set is small and its value does not predict the size of
\chiMF. That the variation of \chiMF\ is so small in the U10 model with
$r_t=18.5$ is a result of the tradeoff between the three mass
functions.    There is no correlation between $\tau$ and \chiMF\ between
data sets since the degree of mass segregation with both time and
position also depends on the IMF and $r_t$.  The detailed reasons for
these trends are unclear, but relate to the more general question
of mass segregation. This matter warrants
further study.

For the U10 model sets  \chiSDP\ is constant  for both of the magnitude
limited SDPs and does not result from a tradeoff between them.  Rather,
the improved behavior can be understood in terms of eq.(\ref{slope}).
For the U10 model sets, the mean mass in the core at the time of best
fit is about 1.0\Msolar, while for the other IMFs it is about
1.2\Msolar. The observed slope of the central SDP is -0.9
(\markcite{Drukier \et 1993}), giving $\zeta_k = 1.9$.  Taking $m_u$ to
be the central mean mass, the  mass for the observable stars which
gives the best match to the SDP is 0.8\Msolar\ for the U10 model sets
and 1.0\Msolar\ for the others.  The rate of change $d\zeta_k\over dm_k$
is larger for the U10 model sets, but the stars of optimal mass reach
the turn off during the 12 to 18 Gyr window. As a result, $\zeta_k$
remains within $\pm 0.1$ of the observed value during the entire
interval.  For the other model sets $\zeta_k$ starts off lower than the
observed value and decreases with time.  Thus the model fits
deteriorate substantially as the age of the model increases.

There are several angles from which to consider the strong age
dependence of the model fits.  Models with poor matches to either the
SDP or the MFs cannot be considered good models of NGC~6397.
Similarly, models with ages significantly different from the isochrone
age of the cluster must also be considered to be in difficulty. Since
the quality of match to the SDP depends on the mass of the
turn-off stars at a given time, the discrepancy could suggest that
there are problems with the stellar modeling.  However, the existence
of the U10 models which do not show the strong time effect, shows that
this is not the case and that model sets having their best models at
the wrong age probably have the wrong IMF.  A somewhat philosophical
question remains.  Is an IMF giving a model set without any age
dependence (such as U10)  to be preferred to one with an
age-dependent quality-of-fit, but with a preferred age consistent with
the isochrone age (such as L05)?  If they are not to be preferred, then
can we use the existence of a preferred age to learn anything about
particular clusters?

\subsection{Robustness}
\label{robustness}
 Up until now the discussion of the matches
between the models and the observations has been conducted at a level
removed from the actual matches.  In this section I will show two
matches in order to bring some meaning to the values and differences in
$\chi^2$.  More such matches are shown in Paper B where the purpose is
to extract information about NGC~6397.

I will begin with a model which has been selected for having the lowest
mean $\chi^2$ of the well-fitting models with  ages between 15 and 17
Gyr.  It is the third best of all the well-fitting models, and has a
mean $\chi^2$ only 0.06 larger than the best model. As well, it happens
to have the lowest \chiSDP\ of all the well-fitting models. This model,
designated t074, is a U10 model, with $W_0=6.00$, $M_0=4.5\times
10^5\Msolar$, $r_t=20.$ pc, and $\rlrt=1.08$.  At the displayed time
(Fig.~\ref{t074}) its age is 15.8 Gyr, $\chiMF=1.54$, and
$\chiSDP=1.06$.  As might be expected, the fit to the SDP is quite
good.  The mass function matches are not as much of a success since
many of the details in the observed MFs are not matched.  Of more
concern, the du Pont:out MF is systematically higher than the model MF
at that radius.  This is probably an effect of the choice of $r_t$. The
other concern is that the model  velocity dispersion is systematically
lower, although still consistent with, the observed velocity dispersion
data.  Overall, this is certainly an acceptable model and demonstrates
the validity of the Fokker-Planck modeling.

By way of comparison, in Fig.~\ref{GG057} I show the well-fitting
model with an age between 15  and 17 Gyr and with the lowest \chiMF. This
is the model with the second best MF fit and is only 0.02 worse in
\chiMF\ than the best model. It is also has the fourth-highest
\chiSDP.  This model, designated GG057, is a U20 model,  with
$W_0=5.35$, $M_0=7.56\times 10^5\Msolar$, $r_t=18.5$ pc, and
$\rlrt=0.78$.  At the displayed time  its age is 17 Gyr, $\chiMF=1.06$,
and $\chiSDP=3.16$. The central part of the SDP is  conspicuously
poor.  Model GG057 also has more stars in the outer region than model t074.
This is not a result of the rescaling of the model SDP (see
\S\ref{comparison}), but is a real effect. That this is so can be seen
in the mass functions; their radial positions in the cluster are
indicated by the vertical lines in the SDP panel.  The MFs for
the low-mass stars are almost identical for both models; the
differences lie for  stars more massive than about 0.3\Msolar.  For all
three MFs, model GG057 has more of these stars than does model t074 and
the size  of the difference increases with radius. 
The velocity data is matched very well.

\section{Discussion}
\label{Discussion}
 I have run eleven sets of models in attempting to
match a  set of observations of the globular cluster NGC~6397.  For any
given IMF and galactocentric distance (parameterized as a fiducial
tidal radius $r_t$) any two of the remaining parameters $W_0$, $M_0$,
and \rlrt\ are independent; the requirement to match all the
observations at the same time fixes the third.  For most of the IMFs
tested the quality of the match to the observations is time dependent
with only a fairly narrow interval in which both the surface density
profile and the mass functions are matched well.  The size of the time
dependencies is determined by the IMF and, to a lesser extent, by
$r_t$. In two model sets with one IMF, only a weak time dependence was seen.
While this may still not be the optimal choice of IMF, as things stand
the existence of preferred times in the other models cannot be used to
constrain the age of NGC~6397 in the face of an IMF without a preferred
age.   What can be said is that  the existence and quality of matching
models can put limits on the existence and numbers of stars with masses
outside the observed range. A small fraction of neutron stars is
required, but not too many. As well, the existence of a very large
number of low mass stars also appears unlikely.  These, and other
constraints relating specifically to NGC~6397, are discussed in more
detail in Paper B.

The generalized rules discussed in \S\ref{guidelines} can be put on a
firmer footing using the results of the various model sets.  The
$\left(W_0,M_0,\rlrt\right)$ surfaces are curved, but rough estimates
can be made of the relationship between changes in the parameters and
the times of best  match to the SDP or the MFs.  Series of models
varying in only one parameter can be taken from the grids and used to
estimate the variation in the times as a function of the parameters.
For the eight models sets shown in Fig.~\ref{grids} the slopes
estimated this way are shown in Table~\ref{grid slopes}. These
numbers are meant to be representative and suggest the range of
variation possible with variations in the IMF. Changes in \rlrt\ are
about three times more effective in changing the time of optimal mass
than the time of core collapse.  Changes in $W_0$ do not effect the
time of core collapse all that much.
The slopes can also be used to quantify the thickness of the
surface.  Since the models are checked intermittently even the
well-fitting models may not have $\Delta t=0$ if they are rerun and
checked more frequently.  From my list of $\Delta t$ values, the most
common one other than zero is 0.4, suggesting that this is the typical
time interval between data saves in the vicinity of well-fitting
models.  The final three columns of Table~\ref{grid slopes} give the
variations in $W_0$, $M_0$, and \rlrt\ which change $\Delta t$ by 0.4
Gyr.  A model with a single one of these parameters changed by the
indicated amount should still be well-fitting.

One thing that is clear from this work is the strong effect the IMF has on
the quality of the model fits.  Additional data can only serve to
further limit the range of initial parameters which match the
observations.  In retrospect, it may have been more fruitful to treat
the three mass functions as independent constraints rather than to use
the combined results.  As a first attempt at such an extensive
comparison, the more limited goal of matching the ensemble of mass
functions simplified the analysis of the results.  The velocity data
does not give as strong a constraint on individual models as does the
SDP and the MFs. The model velocity profiles are much the same for all
well-fitting models in a model set.  The velocities do provide a
stronger limit on the IMF, but the present data set is not precise
enough to make firm statements.

These models still do not include all the effects that are expected to
affect globular clusters.  One such effect is disk shocking. Weinberg
(1994) has included this in his Fokker-Planck code and shown that in
the inner part of the galaxy, clusters can lose substantial amounts of
mass through disk shocking.  This has much the same effect as stellar
evolution mass loss, but takes place for the entire lifetime of the
cluster not just the initial Gyr. The extra mass loss would allow for
models with higher initial masses to reach core collapse at the
present, but also requires high initial concentrations to prevent them
from disrupting entirely.  In Fig.~\ref{rlrt}, only the models on the
right edge of the diagram (those with $W_0>6.5$) would survive based on
Weinberg's preliminary results.  For
more distant clusters the initial concentration can be lower.

The techniques I have used here could certainly be extended to other
clusters, provided sufficiently detailed sets of observation
exists.  Once an IMF which
can give an good match to those observed has been found, and I have
not addressed this question here, it should be fairly straightforward
to locate the range of good models, assuming they exist.  It is
difficult to give explicit rules, even those as rough as the ones given
in \S\ref{guidelines}, which would apply in all cases, but a little
experience with a given data set soon provides these.  The surfaces of
well-fitting models for the last model sets calculated were located
much more quickly than  the first ones.  No hunting phase was
required. One technique is to take a single cut, varying only one of
$W_0$, $M_0$, or \rlrt. Once the $\Delta t=0$ point has been found,
simple extrapolations along a regular grid, following the rules in
\S\ref{guidelines}, quickly find additional extrapolations.  After
several intersection points have been located, estimates can be made
of the shape of the surface. Farther-range extrapolations are often
quite successful and only a minimum of non-useful models need
to be run. Until this has been tested on other data sets, it is
impossible to say how  universally this will apply. For NGC~6397, at
least, these Fokker-Planck models have been quite successful in
matching the observations. The information which can be extracted from
these matches is the subject of the accompanying Paper B.

\acknowledgments

Thanks go to R. Elson for suggesting the contour plots.  This work
was supported by NSERC of Canada and  PPARC of the U.K.
\appendix
\section{A smooth tidal boundary}
\label{app:A}
Assume that the cluster moves in a circular orbit at a distance $R_G$ from
the center of the galaxy and that the mass of the galaxy within $R_G$ is $M_G$.
{}From \markcite{Lee \& Ostriker (1987)} the tidal stripping  is given by
\begin{equation}
{{\partial f(E,t)}\over{\partial t}} = -C_t f(E,t) b(E) t_t^{-1},
\end{equation}
where the stripping rate $b(E)$ is given by
\begin{equation}
    b(E) = \cases{ [1-(E/E_t)^3]^{1/2}  &$E < E_t$, \cr
                   0. & $E \ge E_t$, \cr}
\label{eq:rates}
\end{equation}
and
\begin{equation}
t_t = {{2\pi}\over{\sqrt{{{4\pi}\over{3}}G\rho_t}}}
\end{equation}
is the orbital periods of the cluster about the galaxy.  The
tidal-energy boundary $E_t$  is given by the potential at the radius
which encloses a mean density equal to the tidal density $\rho_t$, a
constant dependent on $R_G$ and the ratio of the initial mass of the
cluster to $M_G$. Note that the potential is defined to be positive
here with $\phi\rightarrow 0$ as $r\rightarrow\infty$. $C_t$ is a
dimensionless constant giving the overall rate of mass loss per orbital
period and is taken to be unity.
The discontinuity in the first derivative of the stripping rate is
obvious in eq. (\ref{eq:rates}).

Let
\begin{equation}
\beta = 1-\left(E\over E_t\right)^3.
\end{equation}
{}From  eq. (\ref{eq:rates}), $b(\beta)=\sqrt{\beta}$ for $\beta>0$ and
is
identically zero for $\beta<0$.  For the region
$\left|\beta\right|\le\epsilon$, where $\epsilon$ is small, I find a
cubic polynomial which has the same values and first derivatives as
$b(\beta)$ at $\beta=\pm\epsilon$.  The required polynomial is
\begin{equation}
b_1(\beta)={\sqrt{\epsilon}\over8}\left[-\left(\beta\over\epsilon\right)^3
          +\left(\beta\over\epsilon\right)^2+5{\beta\over\epsilon}+3\right].
\label{eq:newrate}
\end{equation}

For  $\left|\beta\right|\le\epsilon$ eq. (\ref{eq:newrate}) was used
instead of eq. (\ref{eq:rates}) with $\epsilon$ chosen such that two or
three of the energy grid points would fall within
$\left|\beta\right|\le\epsilon$.

\section{Iterative scheme to ensure self-consistency of the tidal boundary}
\label{app:B}

Following the completion of the Fokker-Planck step, the distribution
function, the phase space functions $p$ and $q$ (qv. \markcite{Cohn
1980}), and the old potential are stored.  I'll refer to this group of
functions as model $\bf F$ and this initial model as $\bf F^0$.  One
iteration of the Poisson solver is done to estimate the tidal radius,
$r_t^0$, and the tidal energy, $E_t^0$. The Poisson solver changes $\bf
F$, so $\bf F$ is reset to $\bf F^0$. Tidal stripping is now done using
$E^0_t$ and a full solution of the Poisson equation is performed
yielding model $\bf F^1$. From $\bf F^1$, $r_t^1$ is calculated. If
$r_t^1=r_t^0$, to sufficient precision,  then model $\bf F^1$ is
self-consistent both between the distribution functions and the
potential (as a result of the Poisson solver) and with respect to the
tidal boundary.  If  $r_t^1\neq r_t^0$, then $\bf F$ is reset to $\bf
F^0$, stripped using $E_t^1$ based on $r_t^1$ and the potential in
model $\bf F^1$ and fed once again through the Poisson solver. This
procedure is repeated until $r_t^n=r_t^{n-1}$ to sufficient precision
(I used a fractional difference of less than $10^{-3}$).

Solving Poisson's equation, even without this iterative scheme, is the most
computer intensive part of the code so it is inefficient to do too many
iterations of the tidal stripping.  If more than one full solution of
Poisson's equation was required to get $r_t$ at a particular time step, then
the next time step was restricted to be no more than one-half the time step
just used. It was found that this procedure worked well and could track the
tidal radius in the late stages of evolution even for very small masses.

\clearpage
\begin{table}
\caption{Stellar lifetimes of stars with $Z=2\times 10^{-4}$. }
\begin{tabular}{lc}
\hline
\hline
Mass &Lifetime \cr
(\Msolar) & (yr)\cr
\hline
         0.8  &     $1.56 \times 10^{10}$\cr
         0.9  &     $1.03 \times 10^{10}$\cr
           1  &     $7.02 \times 10^9$\cr
        1.25  &     $3.23 \times 10^9$\cr
         1.5  &     $1.97 \times 10^9$\cr
         1.7  &     $1.41 \times 10^9$\cr
           2  &     $1.01  \times 10^9$\cr
         2.5  &   $ 5.70 \times 10^8$\cr
           3  &    $3.38 \times 10^8$\cr
           4  &     $1.63 \times 10^8$\cr
           5  &  $  9.98 \times 10^7$\cr
           7  &   $4.99 \times 10^7$\cr
           9  &   $3.15 \times 10^7$\cr
          12  &   $1.98 \times 10^7$\cr
          15  &   $1.45 \times 10^7$\cr
          20  &   $1.02 \times 10^7$\cr
          25  & $ 7.84 \times 10^6$\cr
          40  &  $5.34 \times 10^6$\cr
\hline
\hline
\end{tabular}
\label{ages}
\end{table}

\begin{table}
\caption{IMF bin masses (in \Msolar) and mass fractions.}
\begin{tabular}{cccccccccc}
\hline
\hline
\multicolumn{2}{c}{L05} & \multicolumn{2}{c}{U10} & \multicolumn{2}{c}{U20} &
\multicolumn{2}{c}{U30} & \multicolumn{2}{c}{X2}\cr
$\overline{m}^i_j$ & $M_j$ & $\overline{m}^i_j$ & $M_j$ &$\overline{m}^i_j$ &
$M_j$ &$\overline{m}^i_j$ & $M_j$ &$\overline{m}^i_j$ & $M_j$ \cr
\hline
0.0825 &   0.264 &  0.113 &  0.0811 &  0.113 &  0.0692 &  0.113 &  0.0635 &
0.113 &  0.0978 \cr
 0.174 &  0.0882 &  0.145 &  0.0716 &  0.145 &  0.0612 &  0.145 &  0.0561 &
0.145 &  0.0864 \cr
 0.262 &  0.0480 &  0.187 &  0.0632 &  0.187 &  0.0540 &  0.187 &  0.0495 &
0.187 &  0.0763 \cr
 0.349 &  0.0312 &  0.240 &  0.0557 &  0.240 &  0.0476 &  0.240 &  0.0437 &
0.240 &  0.0673 \cr
 0.436 &  0.0236 &  0.308 &  0.0492 &  0.308 &  0.0420 &  0.308 &  0.0386 &
0.308 &  0.0594 \cr
 0.523 &  0.0200 &  0.396 &  0.0441 &  0.396 &  0.0376 &  0.396 &  0.0345 &
0.396 &  0.0532 \cr
 0.609 &  0.0175 &  0.508 &  0.0442 &  0.508 &  0.0377 &  0.508 &  0.0346 &
0.508 &  0.0533 \cr
 0.696 &  0.0155 &  0.653 &  0.0453 &  0.653 &  0.0387 &  0.653 &  0.0355 &
0.653 &  0.0547 \cr
 0.750 & 0.00353 &  0.750 & 0.00514 &  0.750 & 0.00439 &  0.750 & 0.00403 &
0.750 & 0.00620 \cr
 0.772 & 0.00354 &  0.772 & 0.00516 &  0.772 & 0.00440 &  0.772 & 0.00404 &
0.772 & 0.00622 \cr
 0.794 & 0.00355 &  0.794 & 0.00517 &  0.794 & 0.00441 &  0.794 & 0.00405 &
0.794 & 0.00624 \cr
 0.816 & 0.00355 &  0.816 & 0.00517 &  0.816 & 0.00442 &  0.816 & 0.00405 &
0.816 & 0.00625 \cr
 0.839 & 0.00356 &  0.839 & 0.00518 &  0.839 & 0.00442 &  0.839 & 0.00406 &
0.839 & 0.00625 \cr
 0.863 & 0.00359 &  0.863 & 0.00522 &  0.863 & 0.00446 &  0.863 & 0.00409 &
0.863 & 0.00630 \cr
 0.887 & 0.00358 &  0.887 & 0.00522 &  0.887 & 0.00446 &  0.887 & 0.00409 &
0.887 & 0.00630 \cr
 0.940 &  0.0111 &  0.940 &  0.0161 &  0.940 &  0.0138 &  0.940 &  0.0126 &
0.940 &  0.0195 \cr
  1.04 &  0.0151 &   1.04 &  0.0220 &   1.04 &  0.0187 &   1.04 &  0.0172 &
1.04 &  0.0265 \cr
  1.21 &  0.0238 &   1.21 &  0.0347 &   1.21 &  0.0296 &   1.21 &  0.0272 &
1.21 &  0.0418 \cr
  1.62 &  0.0568 &   1.62 &  0.0828 &   1.62 &  0.0707 &   1.62 &  0.0648 &
1.62 &  0.0999 \cr
  2.83 &  0.0998 &   2.53 &  0.0975 &   2.83 &   0.124 &   2.83 &   0.114 &
2.83 &   0.122 \cr
  4.90 &  0.0617 &   3.75 &  0.0687 &   4.90 &  0.0767 &   4.90 &  0.0704 &
4.90 &  0.0408 \cr
  6.93 &  0.0453 &   4.96 &  0.0535 &   6.93 &  0.0564 &   6.93 &  0.0517 &
6.93 &  0.0204 \cr
  9.80 &  0.0661 &   6.17 &  0.0440 &   9.80 &  0.0823 &   11.1 &   0.122 &
9.80 &  0.0204 \cr
  13.9 &  0.0486 &   7.38 &  0.0375 &   13.9 &  0.0604 &   18.6 &  0.0776 &
13.9 &  0.0102 \cr
  17.9 &  0.0386 &   8.94 &  0.0525 &   17.9 &  0.0481 &   26.1 &  0.0575 &
17.9 & 0.00611 \cr
\hline
\hline
\end{tabular}
\label{IMFs}
\end{table}

\begin{table}
\caption{Representative relations between the time of best match and the model
parameters.
}
\begin{tabular}{llccccccccc}
\hline\hline
\multicolumn{2}{c}{Model set\tablenotemark{a}}  &    $d t_{MF}\over dW_0$ &
$dt_{MF}\over dM_0$\tablenotemark{b} & $dt_{MF}\over d(\rlrt)$ &$d t_{SDP}\over
dW_0$ & $dt_{SDP}\over dM_0$\tablenotemark{b} & $dt_{SDP}\over d(\rlrt)$
&$\Delta W_0$\tablenotemark{c} & $\Delta M_0$\tablenotemark{b,c}&$\Delta
\rlrt$\tablenotemark{c}\cr
\hline
U30 & 18.5  & 4.4 & 2.3 & -39 & 0.0 & 1.8 & -12 & 0.09& 0.8 & 0.01\cr
U20 & 17.   & 4.2 & 2.3 & -30 & 0.6 & 1.8 & -9  & 0.1 & 0.8 & 0.02\cr
U20 & 18.5  & 6.0 & 2.7 & -34 & 1.0 & 1.9 & -12 & 0.08& 0.6 & 0.02\cr
U20 & 20.   & 6.5 & 2.8 & -36 & 2.8 & 2.1 & -15 & 0.09& 0.6 & 0.02\cr
U20 & 21.   & 5.1 & 2.9 & -27 & 0.7 & 2.3 & -7  & 0.09& 0.7 & 0.02\cr
L05 & 18.5  & 7.2 & 3.0 & -38 & 0.9 & 2.2 & -13 & 0.06& 0.4 & 0.02\cr
U10 & 18.5  & 5.9 & 4.0 & -34 & 1.4 & 2.7 & -15 & 0.09& 0.3 & 0.02\cr
U10 & 20.   & 4.0 & 4.3 & -34 & -0.7 & 3.0 & -7 & 0.09& 0.3 & 0.02\cr
\hline\hline
\end{tabular}
\tablenotetext{a}{Stellar data and $r_t$ in parsecs.}
\tablenotetext{b}{Mass measured in units of $10^5\Msolar$.}
\tablenotetext{c}{Change in parameter required to change $\Delta t$ by 0.4
Gyr.}
\label{grid slopes}
\end{table}

\clearpage
\begin{figure}
\caption{Plot of \chiMF\ (dashed) and \chiSDP\ (solid) against model age. The
vertical
dash marks the time of core collapse. The time of optimal mass is at about 14.5
Gyr. This is not a well-fitting model since
the times of minima for the two $\chi^2$ statistics are not coincident.
The discontinuities occur at the times of the data saves.  }
\label{chi curves}
\end{figure}

\begin{figure}
\caption{Contour diagram of \rlrt\ plotted against $W_0$
and $M_0$ for the U20, $r_t=18.5$pc model set.  The contours are based
on the estimates of the position of the $\Delta t=0$ surface indicated
by the squares. The circled squares are well-fitting models.  The dots
outside the contoured region indicate the size of the grid used to
construct the contours. Features on this scale and smaller should be
ignored. The contours are spaced every 0.05, with heavier contours
every 0.25.  The $\rlrt=0.5$ and $\rlrt=1.0$ contours are labeled.
Model GG057
(see \S\protect{\ref{robustness}} and Fig.~\protect{\ref{GG057}})
is the circled square
at $(W_0,M_0)=(5.35,7.56)$.}
\label{rlrt}
\end{figure}

\begin{figure}
\caption{As Fig.~\protect{\ref{rlrt}} but for the ages of the models on the
$\Delta t=0$ surface. The contours are spaced every Gyr and the 12 and
15 Gyr contours are labeled. Time increase from lower-right to
upper-left.  The age range for results to be included in the figures is
10.5 to 19 Gyr and the isochrone age of NGC~6397 is 16$\pm$2.5 Gyr.
For clarity, the positions of the model estimates have not been repeated
and the non-contoured region has been left blank.}
\label{time}
\end{figure}

\begin{figure}
\caption{As Fig.~\protect{\ref{time}} for (a) \chiMF, (b) \chiSDP, and
(c) $(\chiMF+\chiSDP)/2$.  In each diagram the contours are spaced by
0.1 and the  heavier contours every 0.5. Some of the heavier contours
are labeled. In (a) \chiMF\ decreases for older models (cf
Fig.~\protect{\ref{time}}) and in (b) \chiSDP\ increases with age. The
models with the best overall $\chi^2$ cluster along the lower edge of
(c) with the valley referred to in the text lying between the
$\chi^2=1.5$ contours.}

\label{chi contour}
\end{figure}

\begin{figure}
\caption{Minimal \chiMF\ (triangles) and \chiSDP\ (circles) vs. time
for the U20 model set
with $r_t=18.5$pc. The open symbols are minima from actual runs, the filled
symbols are estimates of the value of $\chi^2$ on the $\Delta t=0$ surface.}
\label{chi vs. t}
\end{figure}

\begin{figure}
\caption{Contour diagrams of the $\Delta t=0$ surface for the U20 model sets
with (a)--(c) $r_t=20$pc and (d)--(f) $r_t=17$pc.
The symbols have the same meaning as in
Fig.~\protect{\ref{rlrt}} except that the dotted line
indicates the boundary of the tested region.
(a) \& (c) \rlrt.  The contours are spaced every 0.05. This corresponds
to Fig.~\protect{\ref{rlrt}}.
(b) \& (e) Age. The contours are spaced every 1 Gyr. This corresponds
to Fig.~\protect{\ref{time}}.
(c) \& (f) Mean $\chi^2$. The contours are spaced every 0.1.  This
corresponds to Fig.~\protect{\ref{chi contour}}c.
The heavy contours are as labeled and the direction of increase is indicated
by the arrow.
}
\label{rt=}
\end{figure}

\begin{figure}
\caption{Minimum \chiMF\ (filled squares) and \chiSDP\ (open squares) vs.
time as estimated on the $\Delta t=0$ surface.  The columns are
labeled by their IMF and the rows by the value in pc of $r_t$  used for the
model set. The U20, $r_t=18.5$ pc panel repeats the solid symbols in
Fig.~\protect{\ref{chi vs. t}}. For the model sets where $\chi^2$ varies with
time, the best models are those lying near the intersection of the two
$\chi^2$ locii. The U10, $r_t=18.5$ pc model set gives the
most consistently good models independent of age.   }
\label{grids}
\end{figure}

\begin{figure}
\caption{As Fig.~\protect{\ref{rt=}} for
(a)--(c) the U30, $r_t=18.5$ pc model set;
(d)--(f) the L05, $r_t=18.5$ pc model set;
(g)--(i) the U10, $r_t=18.5$ pc model set; and
(j)--(l) the U10, $r_t=20.$ pc model set. Model t074
(Fig.~\protect{\ref{t074}}) is the circled square at
$(W_0,M_0)=(6.0,4.5)$ in panel (j).  The contour spacing is as in
Fig.~\protect{\ref{rt=}}. The heavy contours have the indicated values.
For each model set, the first panel corresponds to
Fig.~\protect{\ref{rlrt}}, the second to Fig.~\protect{\ref{time}} and
the third to Fig.~\protect{\ref{chi contour}}c. The plotted functions
increase in the direction of the arrow except that in (f) the
$\chi^2=1.9$ contours surround regions of minima with $\chi^2$
increasing towards the outer edge of the contoured area, and in (i)
$\chi^2$  is constant over most of the contoured region, increasing
slightly at the lower boundary.
}
\label{LRTt}
\end{figure}

\begin{figure}
\caption{This figure shows the values of $\chi^2$  vs. time on the
$\Delta t = 0$ surface for each of the three observed  mass functions
and their mean (\chiMF).  The model sets are the same as in
Fig.~\protect{\ref{grids}}. One can see in the U20 column systematic
trends with $r_t$ and these are repeated qualitatively for the two U10
model sets.  The details of the behavior of the matches to the observed
mass functions are a strong function of the IMF.  }
\label{MF lines}
\end{figure}

\begin{figure}
\caption{Comparison between the NGC~6397 observations and model t074 in
the U10, $r_t=20$ pc model set. Clockwise from upper left: The surface
density profile; the FRST mass function; (top) the du Pont:if mass
function and (bottom) the du Pont:out mass function; and the velocity
dispersion profile.  The dashed line in the mass function panels
indicates the shape of the IMF.}
\label{t074}
\end{figure}

\begin{figure}
\caption{As Fig.~\protect{\ref{t074}} for model GG057. The dotted lines
in the SDP panel (upper left) indicate the radial positions of the
three MFs; in  order of increasing distance from the cluster center:du
Pont:if, FRST, and du Pont:out.}
\label{GG057}
\end{figure}

\vfill
\end{document}